# Wolter-I-like X-ray telescope structure using one conical mirror and one quadric mirror

**Sheng-Hao Chen**（陈晟昊）**, Shuang Ma**（马爽）**, Zhan-Shan Wang**（王占山）[1)]

Key Laboratory of Advanced Micro-Structure Materials, Ministry of Education, Shanghai, 200092, China，Institute of Precision Optical Engineering, School of Physics Science and Engineering, Tongji University, Shanghai 200092, China

**Abstract:** Nested multilayer mirrors are commonly used in X-ray telescope structures to increase the collecting area. To balance the difficulty and cost of producing these mirrors, the classical Wolter-I structure has previously been replaced with a conical Wolter-I structure, but it can lead to significantly poorer angular resolution. In this paper, we consider changing one of the mirror shapes (paraboloid or hyperboloid) of the Wolter-I structure to a conical mirror shape, while the other mirror shape remains a quadric surface-type structure, which could thus ensure the imaging quality. Based on a theoretical analysis and the results of simulations, the cone-hyperboloid and paraboloid-cone structures are found to provide the best angular resolutions, and the cone-hyperboloid structure is shorter than the paraboloid-cone structure at the focal length. The cone-hyperboloid structure is then nested to obtain the best on-axis angular resolution and off-axis images.

**Keywords**: cone-quadric structures, quadric-cone structures, nested structure, angular resolution.

**PACS**：95.55.Ka

1） E-mail: wangzs@tongji.edu.cn

## 1. Introduction

X-ray astronomy is a branch of astronomy related to the observation of X-rays from astronomical objects. The grazing incidence X-ray telescope is an important X-ray observation device that has already produced numerous research results that have raised scientific awareness of the unknown universe.

Grazing incidence X-ray telescope structures were pioneered by Wolter[1] in 1952, when he introduced a paraboloid-hyperboloid type 1 (Wolter-I) structure that consisted of a paraboloidal primary mirror and a confocal hyperboloidal secondary mirror. To increase the collecting area, Van Speybroeck[2] proposed the concept of a multilayer nested telescope in 1972. The nested Wolter-I X-ray telescope is an important observation device with very high angular resolution that is mounted on satellites, such as the Chandra X-ray observatory[3–5] and the XMM-Newton.[6–8] This type of telescope can also focus on geometrical collection area, but the fabrication of the mirrors with quadric surface required is both highly difficult and very costly.

On the basis of the Wolter-I structure, several different mirror shapes have been optimally designed for different purposes in the past. To ensure strict satisfaction of the Abbe sine condition, Wolter[9] proposed the Wolter-Schwarzschild structure in his second paper, which exactly fulfilled the Abbe sine condition and thus eliminated the coma aberration for paraxial rays. Werner[10] designed several polynomial with a factor between 2 and 4 of X-ray telescope structures that were appropriated to improve the angle resolution. Conconi[11] and Burrows[12] determined merit functions that were used to optimize the polynomial for large-field X-ray imaging. Harvey[13] designed a hyperboloid-hyperboloid (HH) telescope to optimize the structure for the resolution at a large field of view. Petre[14] and Serlemitsos[15] designed cone-cone type I telescope structures, which is called conical Wolter-I telescope, and is widely used in actual applications, such as in Suzaku[16-18], NuSTAR[19-21] and Astro-H[22-24]. Saha[25] simply added a second-order axial sag to optimize the on-axis image spot of the conical Wolter-I telescope. The conical Wolter-I telescope has a simple principle, where it uses two conical mirrors to replace the paraboloidal and hyperboloidal surfaces; the most important advantage of this is that the difficulty and the cost of fabricating the mirrors both decrease dramatically, and thus the geometrical collection area can be increased by addition of more nested layers. However, the most serious problem with this structure is that the angular resolution is significantly worse.

To obtain either better angular resolution or a better geometrical collection area, researchers have previously made the mirror shapes for both the primary and secondary mirrors with equal levels of optimization. However, these designs have not been adopted in practical applications, except in the conical Wolter-I telescope, because the complicated mirror shapes make them difficult to fabricate and test. An X-ray telescope structure should be designed with equal importance being given to increasing the geometrical collection area while ensuring the best possible angular resolution. At the same time, the difficulty, time and cost of fabrication of the mirrors must also be considered.

In this paper, we consider a change in one of the mirror shapes of the Wolter-I structure to a conical mirror, which could reduce the production cost in comparison to that required for a paraboloid mirror or hyperboloid mirror. The other mirror shape would still have a quadric-type surface, which should ensure the quality of the imaging. First, while keeping the primary mirror as a cone, we optimize the second mirror using three different mirror types: cone-hyperboloid (CH), cone-paraboloid (CP) and cone-ellipse (CE). Second, we keep the secondary mirror as a cone and then optimize the primary mirror using the above three mirror types. After comparison of the theoretical and simulation results for these structures, the optimum structure can then be determined. Finally, we analyze the angular resolution, the field of view and the geometric collecting area of the nested optimum structure in detail.

## 2. Design of cone-quadric structures

CH, CP, and CE mirrors are designed in this section. The positions of the mirrors can be calculated based on the Wolter-I telescope structure that is shown in a schematic diagram in Fig. 1.[26]

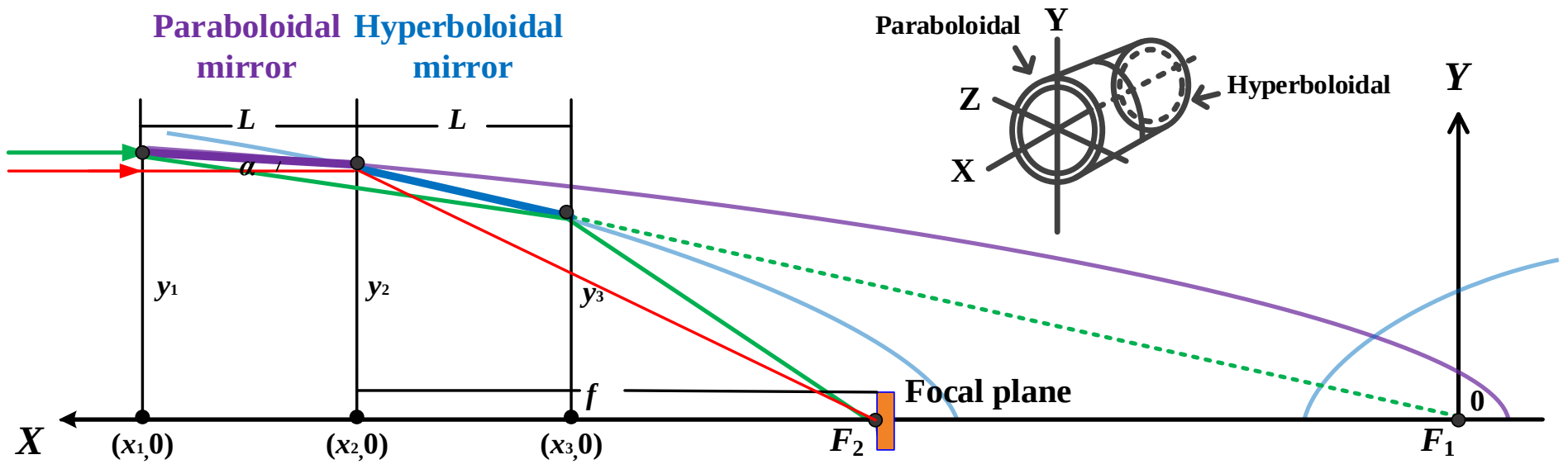

**Fig. 1** Schematic diagram of Wolter-I telescope used in the calculations.

$F_1$ is the focus of the paraboloid and the hyperboloid, and $F_2$ is the other focus of the hyperboloid. The coordinate system is constructed with the origin at $F_1$. The vertical plane at $F_2$ is set as the focal plane of the Wolter-I telescope. $f$ is the focal length of the telescope. $L$ is the axial length of each mirror. $(x_1,y_1)$ and $(x_2,y_2)$ are the coordinates that determine the position of the primary mirror. $(x_2,y_2)$ and $(x_3,y_3)$ are the coordinates that determine the position of the secondary mirror.

A Wolter-I telescope can be completely defined based on three independent parameters: the paraboloid constant $p$, and the two hyperboloid constants $a$ and $b$. The parabolic and hyperbolic equations are set on the basis of the structure in Fig. 1 as:

$$y^2 = p(2x + p) \tag{1}$$

$$\frac{(x-c)^2}{a^2} - \frac{y^2}{b^2} = 1 \qquad c^2 = a^2 + b \tag{2}$$

The initial structural parameters of the telescope are the same as those of the XTP[27]

(X-ray timing and polarization) satellite; the XTP is about to make highly sensitive temporal and polarization observations with good energy resolution in the 1–30 keV range. The telescope parameters are: focal length $f$ of 4550 mm, telescope radius $y_2$ of 225 mm, and the axial length $L$ and the thickness $t$ of each mirror are 100 mm and 0.3 mm, respectively, in both. Table 1 shows the positions of the Wolter-I mirrors in Fig. 1.

**Table 1** Parameters of the Wolter-I structure.

| $\alpha$ (°) | $x_1$ (mm) | $x_2$ (mm) | $x_3$ (mm) | $y_1$ (mm) | $y_2$ (mm) | $y_3$ (mm) |
|---|---|---|---|---|---|---|
| 0.7083 | 9198.1455 | 9098.1455 | 8998.1455 | 226.2329 | 225.0000 | 221.3322 |

We used the same coordinate system as that shown in Fig. 1 to calculate the cone-quadric structure parameters. From Table 1, $\alpha$ is the angle of incidence of the rays on the surfaces are to be equal at the circle of intersection, $(x_1,y_1)$ and $(x_2,y_2)$ can be used to determine the position of the cone mirror, and the position of the quadric mirror is associated with $(x_2,y_2)$ and $(x_3,y_3)$. By setting the telescope focus $F_2$ in Fig. 1 to be the focus of the quadric surface of the secondary mirrors, the functions of the three structures in this coordinate system can be described as follows.

Hyperboloid $$\frac{(x-c_{sh})^2}{a_{sh}^{\ 2}}-\frac{y^2}{b_{sh}^{\ 2}}=1 \quad (3)$$

Paraboloid $$y^2=2p_s(x-x_2+f+\frac{p_s}{2}) \quad (4)$$

Ellipse $$\frac{(x-x_2+f-c_{se})^2}{a_{se}^{\ 2}}+\frac{y^2}{b_{se}^{\ 2}}=1 \quad (5)$$

$a_{\rm sh}$, $b_{\rm sh}$ and $c_{\rm sh}$ are the constants of the hyperboloid of the secondary mirror. $p_{\rm s}$ is the constant of the paraboloid of the secondary mirror. $a_{\rm se}$, $b_{\rm se}$ and $c_{\rm se}$ are the constants of the ellipse of the secondary mirror.

When the rays of parallel to the optical axis are incident on the surface of the primary mirror (cone), the angles of incidence are the same with respect to the slope of the cone mirror $k_c$. The identical angles between all of the reflected light beams and the optical axis are then twice the slope of the cone mirror at $2k_c$. However, the angles between the secondary light beams reflected by the quadric mirrors and the optical axis are quite different because of the different slopes at each point on the secondary mirrors. Figure 2 shows the optical path diagram for beams that are reflected by the quadric mirror.

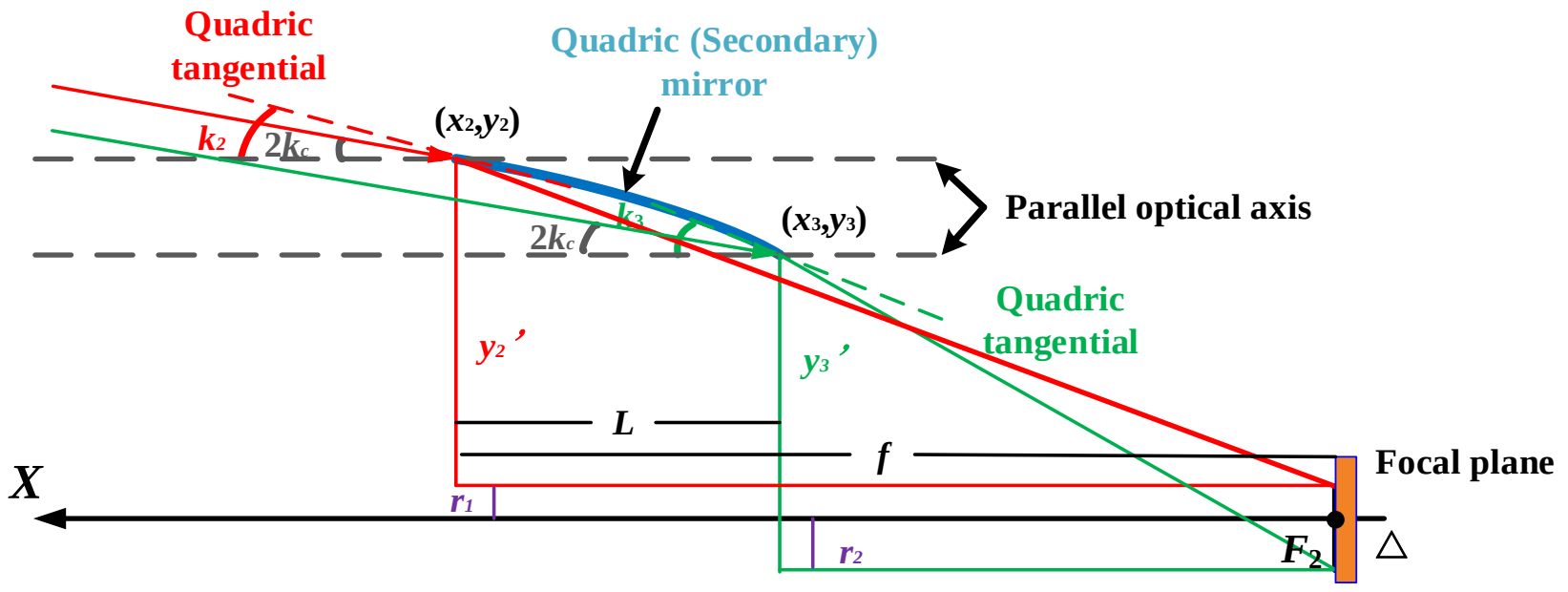

**Fig. 2** Paths of X-rays when reflected by the quadric mirror.

$\Delta$ is the spot size at the focal plane. When the values of $y_1$, $y_2$, and $y_3$ remain constant, the slope of the cone mirror is:

$$k_c = (y_1 - y_2)/L \tag{6}$$

The slopes $k_n$ ($k_2$ and $k_3$) of the quadric mirrors are given by:

$$\text{Hyperboloid} \quad k_n = \frac{(x_n - c_{sh})}{y_n} \cdot \frac{b_{sh}^{\ 2}}{a_{sh}^{\ 2}} \tag{7}$$

$$\text{Paraboloid} \quad k_n = p_s / y_n \tag{8}$$

$$\text{Ellipse} \quad k_n = -\frac{(x_n - x_2 + f - c_{se})}{y_n} \cdot \frac{b_{se}^{\ 2}}{a_{se}^{\ 2}} \tag{9}$$

In the cone-quadric structure, $r_1$ and $r_2$ are the actual vertical heights between the light spots and the optical axis at the focal plane.

$$r_1 = y_2 - y_2' = y_2 - f \cdot (2k_2 - 2k_c) \tag{10}$$

$$r_2 = y_3 - y_3' = y_3 - (f - L) \cdot (2k_3 - 2k_c) \tag{11}$$

Therefore, the value of the radius of spot $\Delta/2$ at the focal plane is the same as the larger of the values of $r_1$ and $r_2$.

Also, the image shown at the focal plane is determined by two conditions:

(1) When $r_1$ and $r_2$ have opposite signs, the image is a spot;

(2) When $r_1$ and $r_2$ have the same sign, the image is a loop.

Equations (6)–(11) have been used to calculate the parameters of these structures and the results are listed in Table 2.

**Table 2** Parameters of CH, CP and CE structures.

| Quadric | $R$ (mm) | $e^2$ | $r_1$ (mm) | $r_2$ (mm) | $\Delta/2$ (mm) |
|---|---|---|---|---|---|
| Hyperboloid | -2.7806 | -1.0012 | 0.5024 | -0.2334 | 0.5024 |
| Paraboloid | -8.1853 | -1 | 6.1520 | 1.9281 | 6.1520 |
| Ellipse | -12.4758 | -0.9982 | 9.7293 | -1.6703 | 9.7293 |

$R$ is the radius of the curve vertex and $e^2$ is the square of the curve eccentricity. Similar results can be obtained via Zemax (a raytracing system) simulations, as shown in Fig. 3.

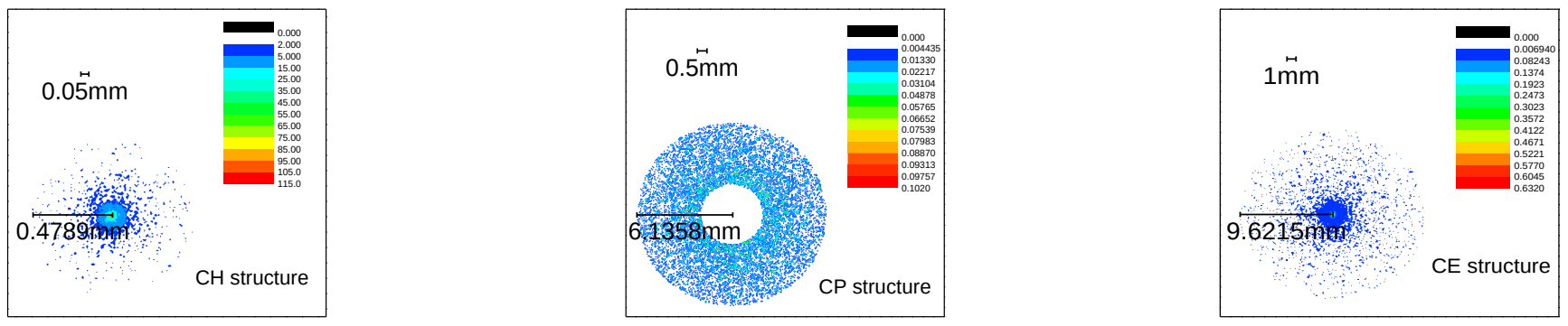

**Fig. 3** Different images of cone-quadric structures produced by Zemax simulations.

Obvious examples of defocusing are shown in Fig. 3. We have optimized the focal lengths of these structures to obtain the best possible angular resolution, and the results are listed in Table 3, as follows.

**Table 3** Optimization of parameters of CH, CP and CE structures.

| Structure | | CH | CP | CE |
|---|---|---|---|---|
| Before optimization | Focal length (mm) | 4550 | 4550 | 4550 |
| | Radius of spot (mm) | 0.4789 | 6.1358 | 9.6215 |
| After optimization | Focal length change (mm) | -3.0614 | +83.8533 | +74.6395 |
| | Radius of spot (mm) | 0.3100 | 2.0896 | 6.1550 |

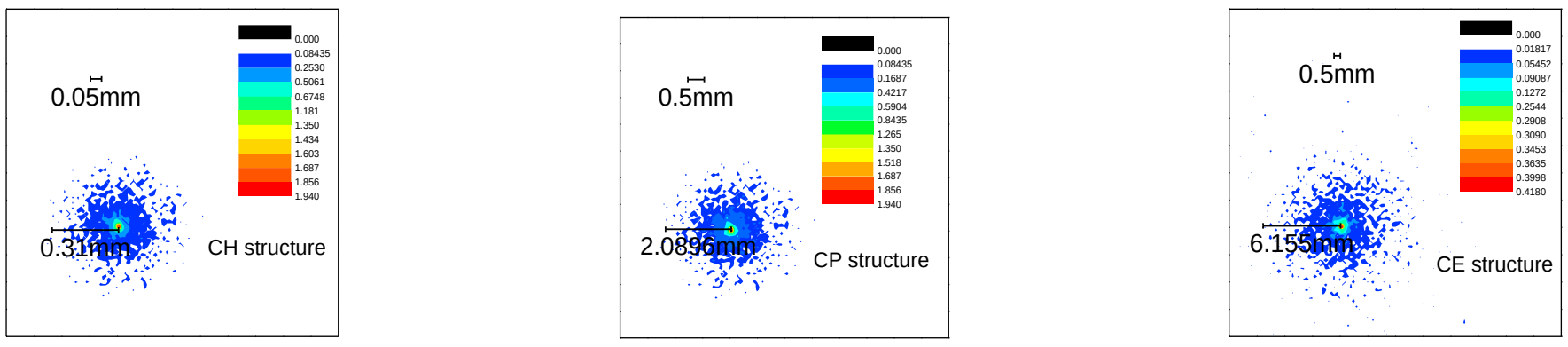

**Fig. 4** Different images of cone-quadric structure after optimization.

Figure 4 shows the different images of the cone-quadric structure after the optimization process. From the calculated results of the simulations, we determined that the CH structure is particularly suitable for engineering applications, because the optimal focus value of this structure is negative, meaning that the telescope has a shorter focal length. This also suggests that the CH structure will provide the best imaging quality.

## 3. Design of quadric-cone structure

The quadric-cone (including PC, HC, and EC) structure is designed in this section. We used the same coordinate system as that shown in Fig. 1, where the telescope focus $F_1$ was set as the focus of the quadric surface of the primary mirrors, and the functions of the three structures in this coordinate system could then be described as follows.

Paraboloid $$y^2 = p_p(2x + p_p \tag{12}$$

Hyperboloid $$\frac{(x+c_{ph})^2}{a_{ph}{}^2} - \frac{y^2}{b_{ph}{}^2} = 1 \tag{13}$$

Ellipse $$\frac{(x-c_{pe})^2}{a_{pe}{}^2} + \frac{y^2}{b_{pe}{}^2} = 1 \tag{14}$$

$p_p$ is the constant of the paraboloid of the primary mirror. $a_{ph}$, $b_{ph}$ and $c_{ph}$ are the constants of the hyperboloid of the primary mirror. $a_{pe}$, $b_{pe}$ and $c_{pe}$ are the constants of the ellipse of the primary mirror. Equations (12)–(14) have been used to calculate these parameters, and the results are listed in Table 4 below.

**Table 4** Parameters of PC, HC and EC structures.

| Quadric | *p* (mm) | *a* (mm) | *b* (mm) | *c* (mm) | $e^2$ | *R* (mm) |
|---|---|---|---|---|---|---|
| Paraboloid | -2.7817 | | | | 1 | 2.7817 |
| Hyperboloid | | 4.9457E+16 | 3.7091E+8 | 4.9457E+16 | 1 | 2.7817 |
| Ellipse | | 4.9457E+16 | -3.7091E+8i | -4.9457E+16 | 1 | 2.7817 |

The results from the hyperboloid and ellipse functions are the same as that for the paraboloid, so we could only simulate the PC structure; the resulting image is shown in Fig. 5.

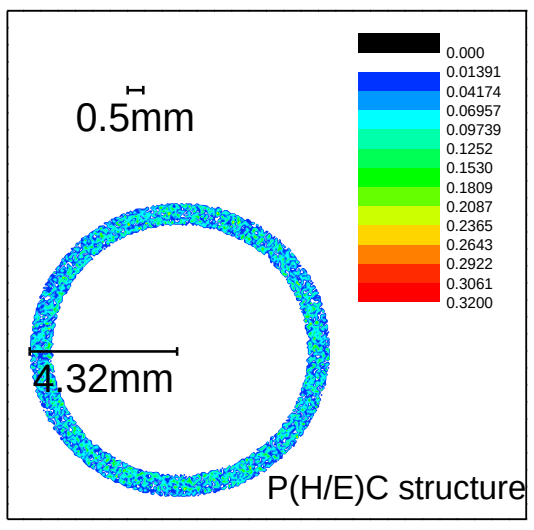

**Fig. 5** Image of quadric-cone.

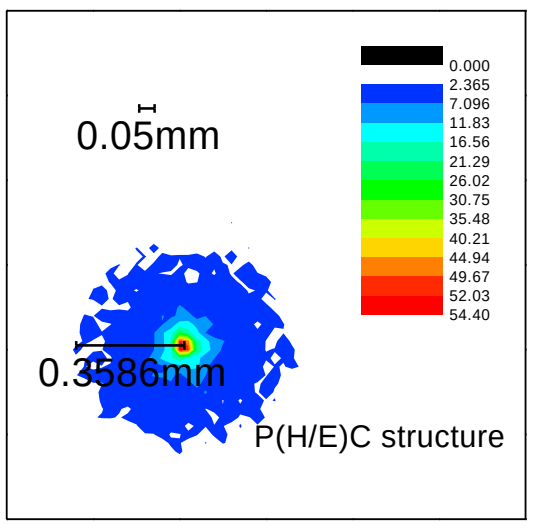

**Fig. 6** Image of quadric-cone after optimization.

Obvious defocusing is also shown in Fig. 5. We have subsequently optimized the focal lengths of the structure to obtain the best possible angular resolution; the resulting parameters are listed in Table 5 and the resulting image is shown in Fig. 6.

**Table 5** Optimization of PC structure.

| Before Optimization | | After Optimization | |
|---|---|---|---|
| Focal length (mm) | 4550 | Focal length change (mm) | +81.0367 |
| Radius of spot (mm) | 4.3215 | Radius of spot (mm) | 0.3586 |

The spot radius of the PC structure is similar to that of the CH structure after optimization. In contrast, however, the optimal focus value of this structure is positive in that the system has a longer focal length. Comparison of the cone-quadric and quadric-cone structures shows that the CH structure produces the best image quality and the optimal focus. The following section gives an analysis of the nested structure.

## 4. Nested structure

To increase the geometrical collection areas to a maximum, the inner surfaces should be sufficiently small to allow all axial rays that strike the next outer surface to pass, and the nested structure is thus designed as shown in Fig. 7[2,28].

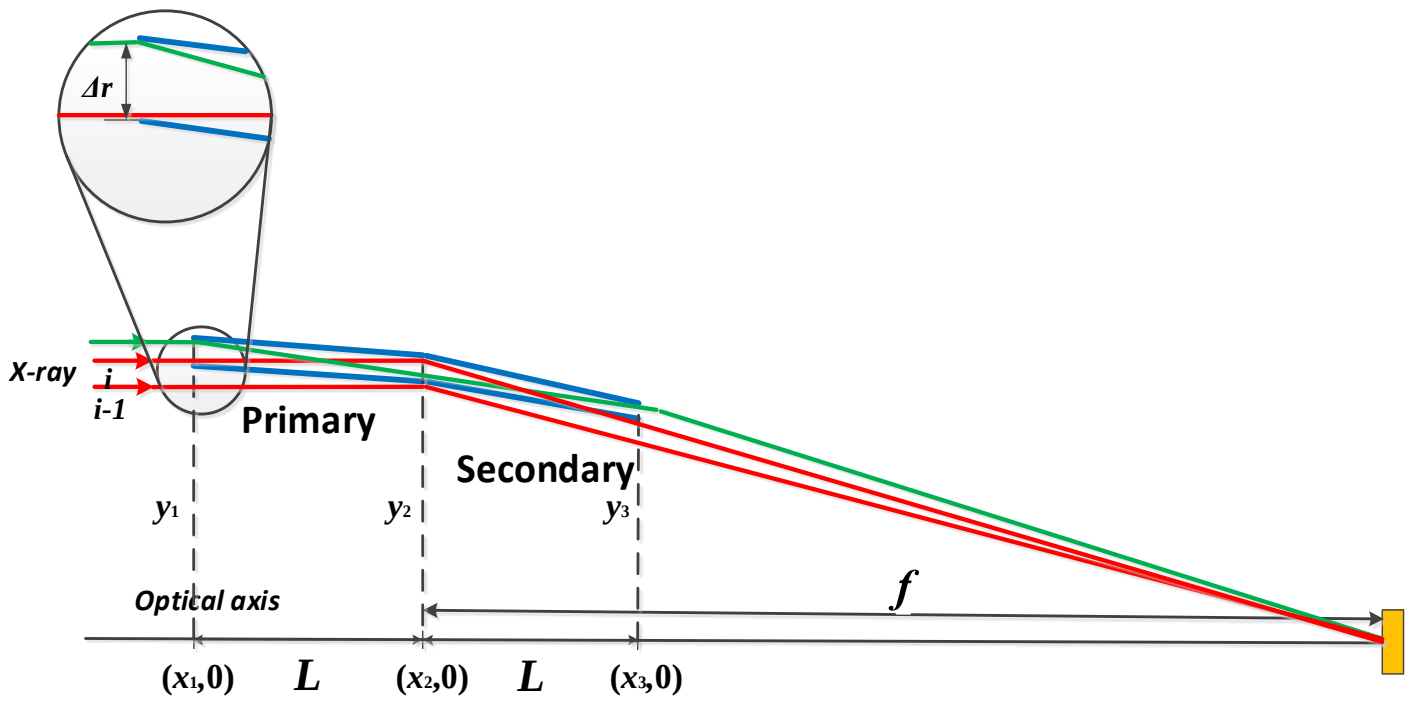

**Fig. 7** Nested structure of the Wolter-I telescope.

The relationship between $y$ and $t$ is given by

$$y_{1,i-1} = y_{2,i} - t \tag{15}$$

When the thickness $t$=0.3 mm of the mirrors is known and $f$=4550 mm, $L$=100 mm, and $y_{2,i}$=225 mm, the positions and the structural parameters of each layer of mirrors can be calculated individually.

A 10-layer nested CH structure is simulated using Zemax, and the resulting image is shown in Fig. 8.

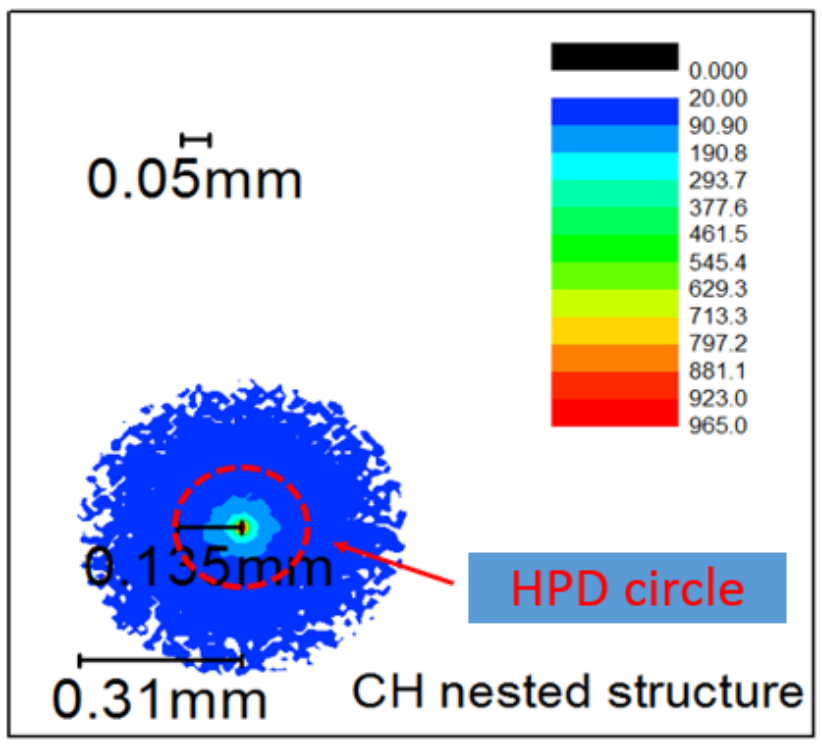

**Fig. 8** Image of CH nested structure.

The central dotted circle denotes the half-power diameter (HPD), and the spot radius of the HPD is approximately 0.135 mm. The angular resolution of the CH structured telescope is thus approximately 12.24 arcsec (HPD).

Under the same nested conditions, the angular resolution of the Wolter-I (PH) structure telescope is approximately 0.10 arcsec (HPD) and the angular resolution of the conical Wolter-I (CC) telescope is approximately 28.58 arcsec (HPD).

The field of view is another important parameter for an X-ray telescope. Table 6 shows the spot diagrams of three types of nested structures on different fields of view, and Fig. 9 shows the changes in resolution (HPD) as the off-axis angles change.

**Table 6** Spot diagrams of three types of nested structures on different fields of view.

| FOV (arcmin) | PH structure | CH structure | CC structure |
|---|---|---|---|

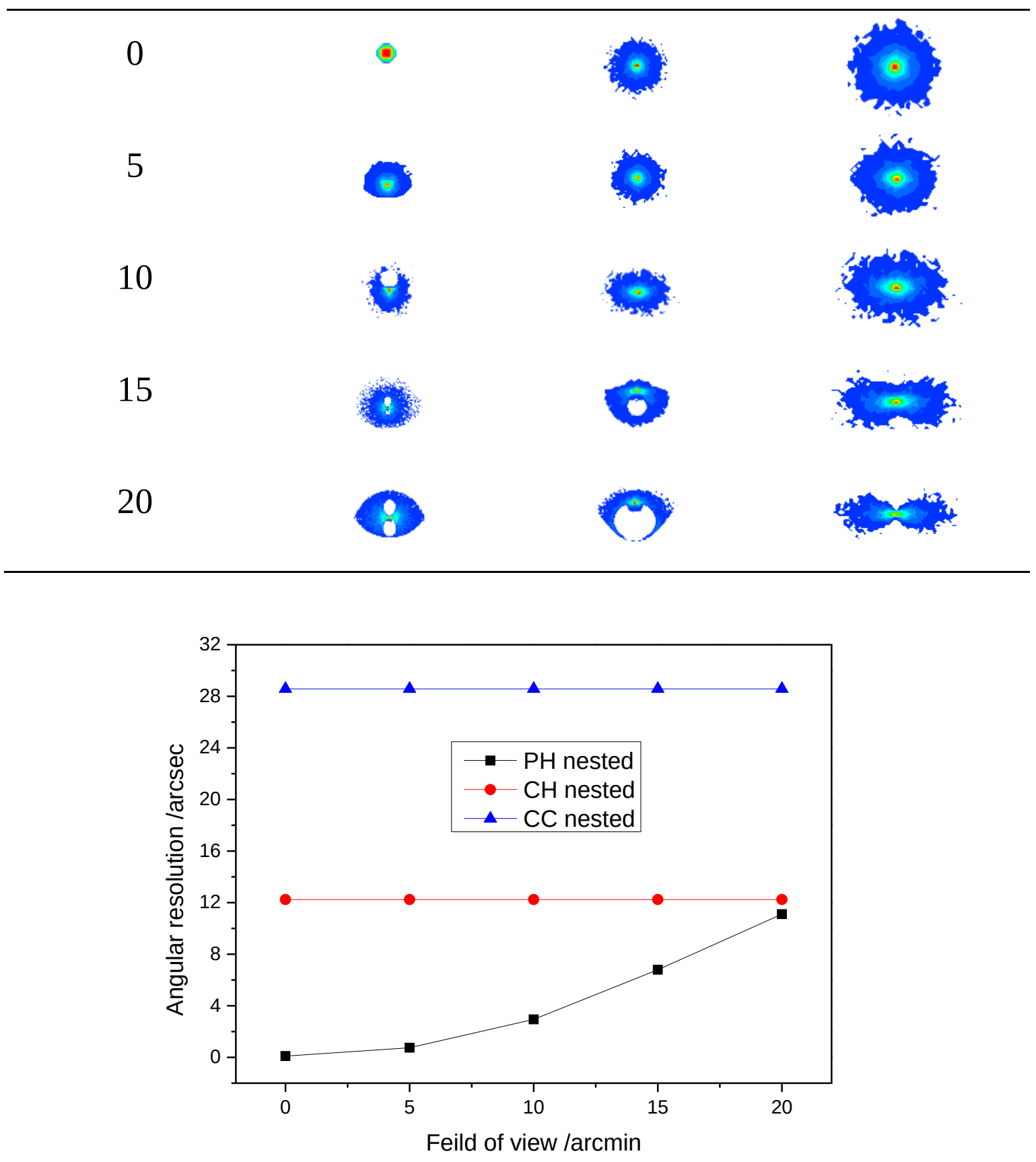

**Fig. 9** Resolution (HPD) changes with change in the field of view.

The angular resolutions of these nested structures are almost the same at the different off-axis angles as that obtained on the optical axis, except for the PH structure.[14,17] In the larger off-axis angle, the spot size of the PH structure is close to that of the CH structure.

Set up the same number of incident rays to the different structures. The rays that are collected from the detector are used to describe the geometrical collection area. Figure 10 shows the change in these rays with the off-axis angle.

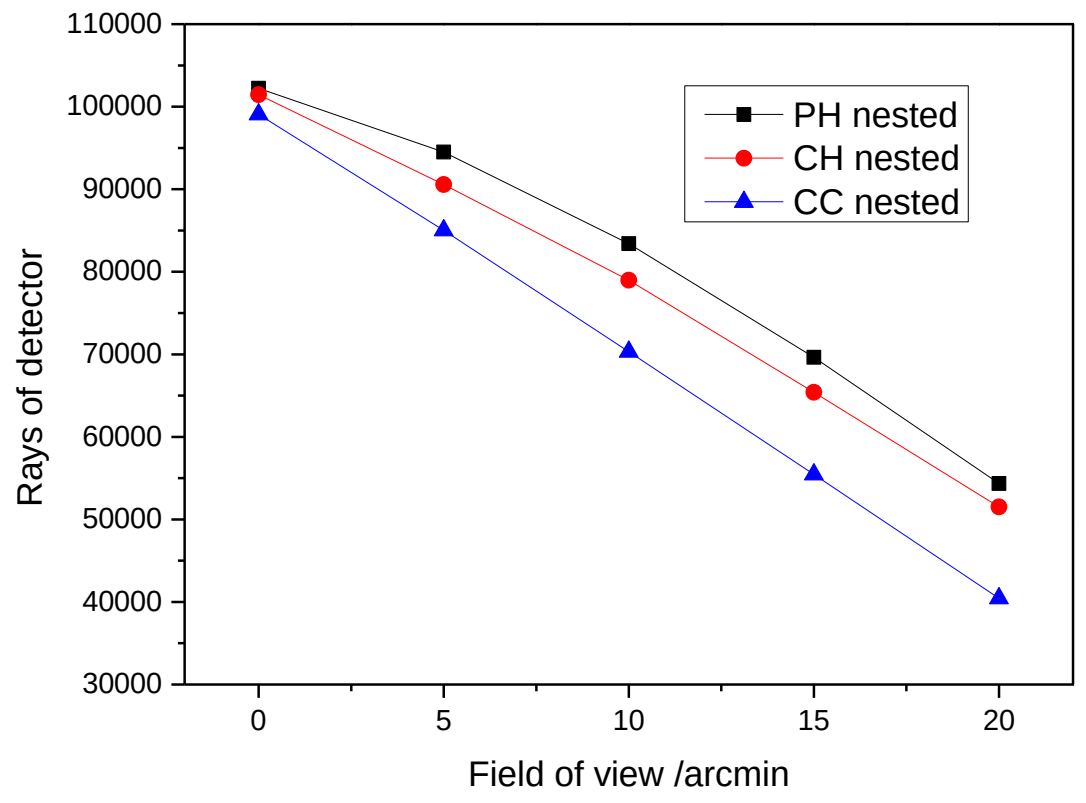

**Fig. 10** Change in rays of detector with change in the field of view.

At the on-axis angle, the geometrical collection areas of the three nested structures are very similar. However, the geometrical collection area of the CH nested structure, which is similar to that of the PH nested structure, is obviously better than that of the CC nested structure at the off-axis angles.

## 5. Conclusions

In this paper, it was first determined that the CH structure, as one of the proposed cone-quadric structures, offers the best imaging quality and a better focal length in them. Second, it was proved that the PC structure was representative of the quadric-cone structures. After comparison and analysis of these structures, the optimum structure was judged to be the CH structure. The results showed that the angular resolution of the CH nested structure is approximately 12.24 arcsec (HPD), and images of the CH nested structure were shown on different fields of view. Therefore, based on consideration of the balance of the angular resolution with the fabrication time and costs, the CH nested structure is selected as a reasonable choice of structure for use in the next stage of our research.

*Acknowledgments*

This work was supported by Supported by CAS XTP project XDA04060605, National Science Instrument and Equipment Development Major Project of Ministry of Science and Technology of China (2012YQ24026402 and 2012YQ04016403).